\documentstyle[sprocl,epsfig]{article}

\def\Journal#1#2#3#4{{#1} {\bf #2}, #3 (#4)}

\def\NPB{{\em Nucl. Phys.} B}
\def\PLB{{\em Phys. Lett.}  B}
\def\PRL{\em Phys. Rev. Lett.}
\def\PRD{{\em Phys. Rev.} D}
\def\ZPC{{\em Z. Phys.} C}

\def\be{\begin{equation}}
\def\ee{\end{equation}}
\def\bea{\begin{eqnarray}}
\def\eea{\end{eqnarray}}

\begin{document}

\title{SOME TOPICS ON DOUBLE HEAVY MESONS:
HEAVY QUARKONIA AND $B_c$ MESON\footnote{A talk 
presented in the OCPA Conference
``Recent Advances and Cross-Century Outlooks in Physics'', 
March 18-20, 1999, at Atlanta}\\
(Advances and outlooks)}

\author{CHAO-HSI CHANG (ZHAO-XI ZHANG)}

\address{Institute of Theoretical Physics, Chinese Academy of Sciences,\\
P.O. Box 2735, Beijing 100080, China\\
E-mail: zhangzx@itp.ac.cn} 


\maketitle\abstracts{The most important advances
achieved recently on the double heavy flavored mesons 
are reviewed. Some problems and outlooks on them are also outlined.}

\vspace{2mm}
\noindent
{\bf 1 Introduction}
\vspace{2mm}

The double heavy mesons, i.e. the heavy
quarkonia $(c\bar c), (b\bar b)$ and 
$B_c$ meson as well as its excited
states, are non-relativistic bound-state.
Their static properties and most decays may be computed 
quite precisely by means of the potential model (PM)
inspired in QCD and relevant theories as well, 
whereas their hadronic production was not
understood so well, although the hadronic production
of $J/\psi$ is so important, not only for perturbative 
QCD (PQCD) theory itself but also for its applications, 
for instance, it is used as a probing tool to detect
specific signature in nuclear processes such as that of 
quark gluon plasma (QGP) etc. Substantial progress on the 
hadronic production has been achieved in recent years. 
Last year, the discovery of the double heavy meson $B_c$ 
opened up new challenges for studying the problems.  
Moreover, at tree level the $B_c$ meson, being a ground state and
in flavor non-singlet, decays by weak interaction only, and
the branching ratios of the decays are sizable\footnote{On 
the contrary, the ground states of heavy quarkonia
(in flavor singlet) decay through 
strong and electromagnetic interactions mainly, that most of the 
branching ratios of the possible weak decays are so tiny beyond the 
capacity of the present experimental techniques, although 
the widths of the weak decays have the 
same order as those of the meson $B_c$.}. Hence 
in addition to the mesons $B$ and $D$, we
will be able to study the weak decays of the heavy flavors 
$b$-quark and $c$-quark with the meson $B_c$. 

\vspace{2mm}
\noindent
{\bf 2 Advances and Problems}
\vspace{2mm}

Fresh progressions on the double heavy mesons are outlined below.

\vspace{2mm}
\noindent
{\em 2.1 Advance on Fragmentation Functions of the Double Heavy Mesons}
\vspace{2mm}

\begin{figure}\begin{center}
   \epsfig{file=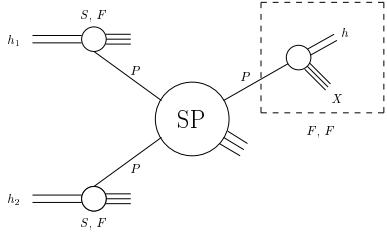, bbllx=137pt,bblly=207pt,bburx=475pt,bbury=423pt,
width=5.5cm,angle=0}
\caption{Factorization of an Inclusive Production for Hadron Collisions
and Definition of Fragmentation Function (F.F.).}
\label{figa}
\end{center}
\end{figure}

It is known that perturbative QCD (PQCD) 
works efficiently with its factorization theorem\cite{twk}.
Namely the `hard part' involved a concerned process
can be calculated perturbatively, but the 
`soft part', e.g. structure functions 
(S.F.s) and fragmentation functions (F.F.s),
being in non-perturbative nature, relating to 
the `color confinement' and independent on the specific
process, are just treated as `universal factors'. 
The factorization theorem and definition of a 
fragmentation function may be illustrated precisely 
by Fig.\ref{figa} and (\ref{ft}) with an inclusive process.
\setcounter{figure}{1}
\bea
\displaystyle d\sigma&=&\sum_{i,j,k}\int dx_1\int dx_2\int dx_3 F_{h_1}^i(x_1,
\mu_F)\cdot F_{h_2}^j(x_2,\mu_F) \nonumber \\
&\cdot& d\sigma_{i,j\rightarrow k,x}(x_1,x_2,
x_3,\mu_F) \cdot D_k^h(x_3,\mu_F), 
\label{ft}
\eea
where $F_{h_1}^i(x_1,\mu_F)$ and $F_{h_2}^j(x_2,\mu_F)$ are the 
S.F.s to depict the probabilities for finding a parton $i$ in the
hadron $h_1$ with momentum fraction $x_1$
and a parton $j$ in the hadron $h_2$ with momentum fraction $x_2$
respectively; 
$d\sigma_{i,j\rightarrow k,x}(x_1,x_2,x_3,\mu_F)$ is the cross-section
corresponding to the `hard subprocess' (the SP-part in Fig.\ref{figa}); 
the factor $D_k^h(x_3,\mu_F)$ is 
just the F.F., which depicts a probability for the 
produced parton $k$ to form a hadron $h$ with a momentum fraction $x_3$
of the parton $k$. Here $\mu_F$, denoting an energy scale, 
is noted where all the factors
in the equation are matched, and 
sometimes it is omited for simplification 
if there is no cofusion.
The fragmentation functions, being considered independent of
the specific process, contain nonperturbative effects,
that generally they are considered not 
to be calculable, but may be determined 
by measurement(s). Thus
F.F.s are important blocks for PQCD calculations.

\begin{figure}\begin{center}
   \epsfig{file=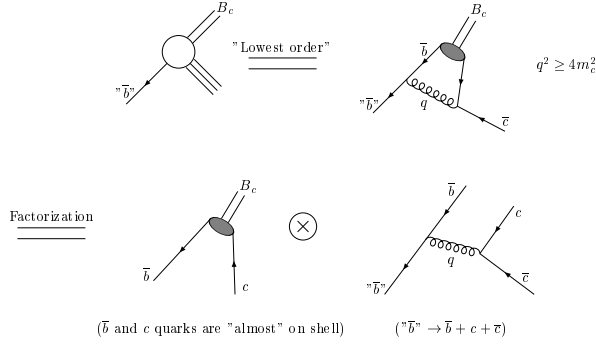, bbllx=126pt,bblly=220pt,bburx=570pt,bbury=496pt,
width=8cm,angle=0}
\caption{The Further Factorization for a Fragmentation Function
of the Double Heavy Meson $B_c$ (an example).}
\label{figb}
\end{center}
\end{figure}

Recently, an advance is to be realized that the fragmentation 
functions (F.F.s) for double heavy mesons can be factorized 
further, and some of them, for the so-called color singlet 
ones, even can be calculated out precisely
without any free parameter\cite{chang1,bra,kis}. 
Taking a F.F. of the meson $B_c$ as an example, let me
with Fig.\ref{figb} to illustrate the `further factorization'. 
Note here that the virtual gluon 
which creates a pair of $c, \bar c$ 
as depicted in Fig.2 is crucial 
for the further factorization. It is because that the gluon
carries a time-like momentum,
and always has $q^2\geq 4m_c^2 \gg \Lambda_{QCD}^2$, so that it
is guaranteed the factor involving the gluon being
perturbative. Namely, due to the very virtual gluon, 
the further factorization makes sense, and we may calculate out 
the perturbative part of the F.F. precisely.
In the cases, when the `residual' factor of the further 
factorization is in color singlet, although being non-perturbative, 
the residual factor may be directly related to the wave 
functions of the potential model directly, i.e. it can be calculated
precisely without any free parameter.
Therefore a fragmentation function
for a double heavy meson can be calculated out 
either totally if the residual factor is in color-singlet 
which may be related to the wave function of potential model directly, 
or partly i.e. only the virtual gluon part may be calculated out and 
the residual factor, as a factor to be determined, 
is left in the `final result' if the residual factor is in color
non-singlet.

To take an example and to see the general features, a
typical fragmentation function, calculated out  
by the method first\cite{chang1}, has the behavior:
\bea
\displaystyle D_{\bar b}^{B_c}(&z&,m_{B_c})\propto
\alpha_s^2(m_{B_c}^2)|\psi(0)|^2\frac{z(1-z)^2}{(a_2z-1)^6}
\cdot\{[2a_1z -3(a_2-a_1)(1-a_2z) \nonumber \\
&\cdot&(2-z)](1-a_2z)z +6(1+a_1z)^2(1-a_2z)-8a_1a_2z^2(1-z)\},
\label{ff}
\eea
where $|\psi(0)|$ is wave function at original
of the $B_c$ meson;
$\alpha_s^2(m_{B_c}^2)$ is the QCD coupling 
constant at $m_{B_c}^2$; and $a_1=m_c/m_{B_c}, a_2=m_b/m_{B_c}$.

\vspace{2mm}
\noindent
{\em 2.2 Advances: `Puzzle' and Solution for Charmonium Production}
\vspace{2mm}

Thanks to the realization about the further factorization
for the fragmentation functions
of double heavy mesons, various 
fragmentation functions not only for the mesons $B_c, B_c^*,
\cdots$ but also for the heavy quarkonia
$(c\bar c)$ and $(b\bar b)$ have been computed. 
For the hadronic charmonium production,
on one hand, the experimental advance is
the data have been accumulated for many years, 
and new achievements in experimental techniques have made it
possible to separate the prompt production 
events of charmomia $J/\psi, \psi', \chi_c, \cdots$ from
those events as decay-products of the
produced $B$ mesons in hadron collisions
successfully\cite{cdf1}; on the other hand, 
the theoretical advance on the 
fragmentation function computation for those of color singlet:
the fragmentation functions have been calculated out precisely 
by means of the wave functions obtained by potential model
(all the parameters appearing in the calculation 
having been determined well elsewhere), therefore thorough 
comparisons between theoretical predictions and experimental data 
become possible and interesting. Indeed the comparisons 
were carried through just after the advances being achieved.

An important result of the comparison, the so-called `$\psi'$ surplus 
puzzle', rose and confirmed. 
Namely, of the prompt production of $J/\psi$ and $\psi'$
in hadron collisions, the experimental data are greater than theoretical 
prediction by one to two orders in magnitude and the behavior of the
$P_T$ (the transverse momentum of the produced mesons) dependence 
for experimental data and theoretical prediction
is also inconsistent, if only the color singlet F.F.
is taken into account. 
We would like to note here that the advance on the computation of the 
fragmentation functions for double heavy mesons is crucial to recognize 
the puzzle, namely, without the advance, the fragmentation functions would 
be able only to be determined by experimental measurement in whole, 
that the deviation (the puzzle) would be `absorbed' into the 
determination to disappear.
Based on analysis of the puzzle, Braaten and Fleming proposed the
so-called color octet mechanism first\cite{oct1}. 
Following them, many groups applied the proposition to various 
processes widely, including to nuclear processes\cite{oct2} to 
explore the consequences of the new mechanism. Since in the
proposition a new undetermined parameter, the so-called
color-octet matrix element (C.O.M.E.) which is of non-perturbative 
feature and uncalculable, is introduced, hence the solution of the 
`surplus puzzle' essentially is to determine the matrix element 
first and then to test it with the other available data. Only
with constant matrix elements determined by fitting data
to obtain correct behavior of the $P_T$ dependence 
of the production has `absolute meaning' for the proposition. 
The reference\cite{oct1} paid attention to the $P_T$ dependence of the 
production on large $P_T$ mainly, but afterwards others tried to have 
a complete determination of the matrix elements for various color 
octet components by fitting whole behavior of $P_T$-dependence 
(at large and small $P_T$)
of the hadronic production\cite{cho,co1,co2}.

\setcounter{figure}{2}
\begin{figure}\begin{center}
   \epsfig{file=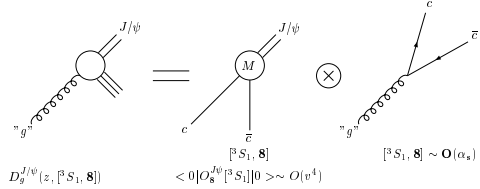, bbllx=100pt,bblly=180pt,bburx=536pt,bbury=376pt,
width=7cm,angle=0}
\caption{1. The Color Octet Component $c\bar c([^3S_1, {\bf 8}])$.}
\label{figc1}
\end{center}
\end{figure}

Taking the charmonium meson $J/\psi$ as an example let me outline 
the key point of the proposition. 
A physical state of $J/\psi$, in general, should be expressed 
in Fock space:
\bea
|J/\psi\rangle&=&O(v^0)|c\bar c([^3S_1 {\bf 1}])\rangle 
+O(v)|c\bar c([^3P_J {\bf 8}])g\rangle +
O(v^2)|c\bar c([^1S_0 {\bf 8}])g\rangle \nonumber\\
&+&O(v^2)|c\bar c([^3S_1 {\bf 8}])gg\rangle 
+O(v^2)|c\bar c([^3D_J {\bf 1,8}])gg\rangle +\cdots, 
\eea
where $v$ is $4$-velocity of the constituents of the meson $J/\psi$,
and, due to $J/\psi$ being heavy nonrelativistic bound state, 
$v^2 \simeq 0.3$ is acceptable; $g$ indicates a `valance' gluon; 
and the factor $O(v^i)$ at the front of each term indicates the
relative order of the term which is obtained
based on the counting rules of the effective
theory: nonrelativistic QCD (NRQCD)\cite{lap}.
It is known that $J/\psi$ is a `flavor singlet' meson, thus it has
the same quantum number in flavor as a gluon, especially, its component 
in the Fock space expansion $(c\bar c[^3S_1 {\bf 8}])$ has the same quantum
numbers of a gluon. As a result of the fact, $J/\psi$
production in hadron collisions will present a quite complicated feature
i.e. in a sense it may be considered as a kind of `mixing' between
the components of higher Fock space with a gluon. 
The fragmentation function of a gluon to $J/\psi$, 
$D^{J/\psi}_g(z,\mu_F=2m_c)$, 
may be used as an example to 
illustrate how to solve the surplus
by the `mixing'. For charmonium, in general, 
we have $v^2\simeq 0.3$ and
$\alpha_s(4m_c^2)\simeq 0.2\sim 0.3$, therefore, there are
components for the fragmentation function, which have contribution 
in the same order of magnitude, 
according to naive NRQCD power-counting rule, are listed below:

\setcounter{figure}{2}
\begin{figure}\begin{center}
\epsfig{file=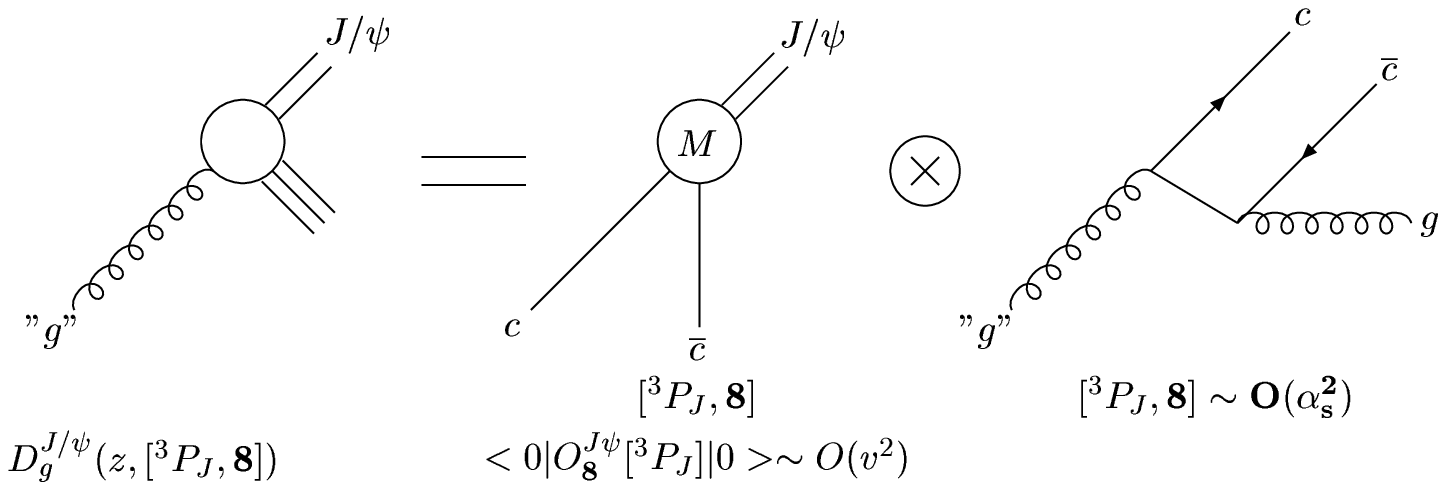, bbllx=100pt,bblly=194pt,bburx=540pt,bbury=344pt,
width=7cm,angle=0}
\caption{2. The Color Octet Component $c \bar c([^3P_J, {\bf 8}])$.}
\end{center}
\end{figure}

{\it 1. The color octet component $(c\bar c[^3S_1, {\bf 8}])$}
({\bf Figure 3:1}):
\be
D^{J/\psi}_g(z,[^3S_1,{\bf 8}])=\langle O^{J/\psi}_8[^3S_1,{\bf 8}]
\rangle \times \Gamma_{g \rightarrow (c\bar c[^3S_1, {\bf 8}])}(z).
\ee

{\it 2. The color octet component $(c\bar c[^3P_J, {\bf 8}])$ 
and $(c\bar c[^1S_0, {\bf 8}])\;$} ({\bf Figure 3:2}):
\be
D^{J/\psi}_g(z,[^3P_J, {\bf 8}])=\langle O^{J/\psi}_8[^3P_J, 
{\bf 8}] \rangle
\times \Gamma_{g\rightarrow (c\bar c[^3P_J, {\bf 8}])g}(z),
\ee
\be
D^{J/\psi}_g(z,[^1S_0, {\bf 8}])=\langle O^{J/\psi}_8[^1S_0, 
{\bf 8}] \rangle
\times \Gamma_{g\rightarrow (c\bar c[^1S_0, {\bf 8}])g}(z).
\ee
Note that for shortening in Fig.3:2, 
only $(c\bar c[^3P_J, {\bf 8}])$ is 
presented, but $(c\bar c[^1S_0, {\bf 8}])$, being similar,
is not.

{\it 3. The color singlet component $(c\bar c[^3S_1, {\bf 1}])$}
({\bf Figure 3:3}):
\be
D^{J/\psi}_g(z, [^3S_1, {\bf 1}]) = \langle O^{J/\psi}_8[^3S_1, 
{\bf 1}] \rangle \times
\Gamma_{g\rightarrow c\bar c([^3S_1, {\bf 1}])gg}(z).
\ee
\setcounter{figure}{2}
\begin{figure}\begin{center}
\epsfig{file=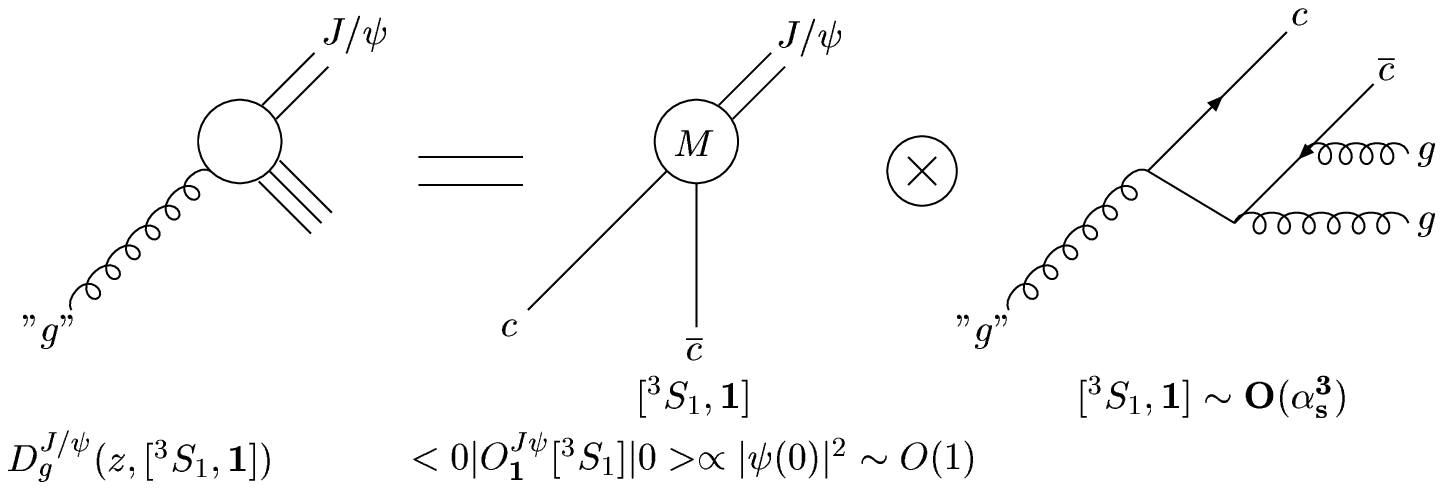, bbllx=100pt,bblly=185pt,bburx=530pt,bbury=340pt,
width=7cm,angle=0}
\caption{3. The Color Singlet Component $c\bar c([^3S_1, {\bf 1}])$.}
\end{center}
\end{figure}
Based on power counting and as indicated in Figs.3:1-3 
no matter the component in the color-octet or in color-singlet,
each may contribute to the F.F. in the same order of
magnitude, that we should treat them `equally'.

\vspace{2mm}
\noindent
{\em 2.3 The Problem(s) of the Proposition}
\vspace{2mm}

Indeed the color octet proposition is very attracting because it is 
on NRQCD formalism and can solve the puzzle 
by choosing the color octet matrix
elements (C.O.M.E.) properly. Whereas the C.O.M.E. cannot be calculated
due to its nonperturbative nature (there is a long way to go also
for Lattice QCD), and furthermore, exactly to say, 
the proposition has not been well-proved yet.

The proof of the proposition should be either 
to show the determination of C.O.M.E. is independent of the 
chosen processes,
or one and more of its characteristic signatures are 
observed in experiments. Whereas the status of this kind of proof now 
is that a lot of suggestions on the characteristic 
signatures are proposed but no one has been observed in experiments 
definitely\footnote{One clean signature of the proposition was
proposed in reference\cite{bch}, but considering to the 
backgrounds pointed out
in Ref.\cite{chq} and experimental errors, 
one cannot conclude a clean signature
has been observed. The other suggestions are worse than the mentioned one.};
for the determination of C.O.M.E., the situation is neither definite:
An example for the situation of the determination of 
C.O.M.E.\cite{cho,co1,co2} is shown in Table 1.

\vspace{0.2cm}
\begin{center}
\centerline{Table 1: Color-octet matrix element linear combinations.}
\vspace{0.1cm}
\footnotesize
\begin{tabular}{|c|c|c|}
\hline
Matrix elements & $\langle O^{j/\psi}_{\bf 8}(^3S_1)\rangle (in\; GeV^3)$
& $\frac{\langle O^{j/\psi}_{\bf 8}(^3P_0)\rangle}{M_c^2}+
\frac{\langle O^{j/\psi}_{\bf 8}(^1S_0)\rangle}{3} (in \; GeV^3)$ \\
\hline
NRQCD scaling order& $M_c^3v^7$  & $M_c^3v^7$ \\
Cho et al\cite{cho} & $(6.6\pm 2.1)\times 10^{-3}$ 
& $(2.2\pm 0.5)\times 10^{-2}$ \\
Beneke et al\cite{co1} & $(1.06\pm 0.14)\times 10^{-2}$ 
& $ \sim $ \\
Sridhar et al\cite{co2} & $(1.26\pm 0.33)\times 10^{-2}$ 
& $(3.14\pm 0.58)\times 10^{-2}$  \\
\hline
\end{tabular}
\end{center}
\vspace{0.2cm}
Note: those of Beneke et al\cite{co1} are taken only
for $\langle k_T \rangle=0$.
The values in the table are obtained only based on the data
of Tevatron, and those from the lepton(photon)-production 
and fixed target ones are not
involved because of more complications. The values of
C.O.M.E. in the table are quite `scattering' 
that I think it is hard to draw any conclusion 
about proving the proposition. 

\vspace{2mm}
\noindent
{\em 2.4 $B_c$ Mesons: Theoretical Predictions 
and Experimental Discovery}
\vspace{2mm}

The meson $B_c$ is the unique one of the possible double heavy mesons
in flavor non-singlet. Being a double heavy meson, 
its properties may be estimated well with the potential model
and recent advances.

Based on the potential model, the estimated mass of the meson is not 
very large in the region $6.2-6.4$ GeV\cite{mas}.

The lifetime of the meson may be estimated by the
spectator mechanism and the `CKM favor' 
annihilation of the two components
$c$-quark and $\bar b$-quark into $c, \bar s$-quarks and $\tau^+,\nu$-leptons
(there is helicity suppression that those
of annihilating into the other
possible quark pairs and lepton pairs besides into $c, \bar s$-quarks 
and $\tau^+,\nu$-leptons may be ignored safely).
The two spectator terms are the same as those for $B$ meson 
and for $D$ meson respectively, thus 
they may relate to the widths of the $B$ meson 
and the $D$ meson i.e. we may estimate the contribution
of the two spectator terms to the width of the meson $B_c$ 
from the experimental measurements
on the lifetimes of the mesons $B$ and $D^0(D_s)$
respectively, and the annihilation contribution may be computed 
easily with the wave function (decay constant) of the meson 
$B_c$. The estimated lifetime of the meson 
$B_c$ is about $0.4\times10^{-12}$s\cite{dec,life,dec1}.

Since the mass of the meson is not very great,
the production of the meson had not been expected 
so `difficult' as realized now: Of the present facilities, for 
experimental detecting only at Tevatron, enough number of
$B_c$ mesons can be produced; in $e^+e^-$ colliders only
at $Z$ resonance LEP-I marginal number of the mesons 
can be produced; of the planned facilities, 
at such a high energy hadronic collider
LHC, numerous mesons will be able to be 
produced\cite{chang1,prod,prod1}.
Quantitative values depend on the cuts of transverse 
momentum $P_T$ and rapidity $y$ quite sensitively. 
Comparative study on the
production calculated by full $\alpha^4$ calculation of PQCD and
by simplified `fragmentation mechanism' showed that due to 
convolution in the 
total cross section the differences of the two approaches
were smoothed out accidentally\cite{prod1}. 

As for the weak decays of the meson $B_c$, 
there are many decay channels that 
contain a charmonium in the final state. 
Especially, when the charmonium in final state is $J/\psi$, 
the decays may be used as a characteristic signature for identifying 
the produced meson $B_c$ because of the detecting advantages of 
$J/\psi$: narrow width and sizable branching ratio to a lepton pair 
$l\bar l$ etc. To compute all the decays, 
there are many ways\cite{dec,dec1,dec2}, but let me pick up one,
the so-called generalized instantaneous approximation
approach\cite{dec,ins} to illustrate a little more in detail.
The non-relativistic wave functions of 
the mesons $B_c$ and $J/\psi$
may be calculated out in potential model framework, 
but there may be a great (relativistic) momentum recoil ($m_{B_c}
\simeq 6.3$ GeV, $m_{J/\psi}\simeq 3.0$ GeV, i.e. the produced $J/\psi$
may move relativistically in C.M.S. of $B_c$). The so-called generalized 
instantaneous approximation approach is to start with a relativistic
formulation, which is based on Bathe-Salpeter 
(B.S.) equation for the double heavy mesons and Mandelstam 
method for the transition matrix elements\cite{mend},  
to write down the relevant decay matrix element 
and then to make an `instantaneous approximation' on 
the whole matrix element in a similar way as the `original 
instantaneous approximation' on B.S. equation done by Salpeter first. 
Finally as the result of the approach, the matrix element 
is related to certain operators sandwiched by 
the nonrelativistic wave functions (the instantaneous `limit'
of the B.S. equations) properly.
Let me quote the semileptonic decay widths estimated
by the approach and some other approaches in Table 2.

\vspace{0.2cm}
\begin{center}
\centerline{Table 2: Exclusive semileptonic decay width for 
various modes}
\vspace{0.15cm}
\footnotesize
\tabcolsep12pt
\begin{tabular}{|c|c|c|c|}
\hline
Mode & Reference\cite{dec} (in $10^{-6}$ eV) & Reference\cite{dec1}
(in $10^{-6}$ eV)  \\
\hline
$B_c \to  \eta_c   + l^+ \nu_l$ &  14.2 & 10.6  \\
$B_c \to  J/\psi   + l^+ \nu_l$ &  34.4 & 38.5 \\
\hline
\end{tabular}
\end{center}
\vspace{0.2cm}

The meson $B_c$ was searched for at LEP-I, but not seen\cite{lep}. 
It was just
discovered by CDF Collaboration 
last year\cite{cdf}. Now no one 
doubts the discovery, although CDF Collaboration was based 
on about 20 events of the cascade decays 
$B_c \rightarrow J/\psi+l\nu, J/\psi \rightarrow \mu^+\mu^-$
and reported three measurements with quite large error bars only. 
The three are its mass, lifetime and a combination of the production cross 
section and the decay branching ratio of the meson.

In summery for comparison, the CDF results of the three measurements:
\be
m_{B_c} = 6.40 \pm 0.39 \pm 0.13, 
\ee
\be
\tau_{B_c}=0.46^{+0.18}_{-0.16}(stat.)\pm0.03(syst.)\times 10^{-12} s,
\ee 
and a combination of the production 
cross-section and
decay branching ratio `normalized' by those of $B$ meson's:
\bea
\displaystyle R & \equiv & \frac{\sigma\cdot Br(B_c\to J/\psi+l^++\cdots)}
{\sigma Br(B^+\to J/\psi+K^+)}\nonumber\\
& = &0.132^{+0.041}_{-0.027}(stat.)\pm 0.031(syst.)
^{+0.032}_{-0.020}(lifetime).
\eea
The ratio $R$, the mass $m_{B_c}$ and the lifetime $\tau_{B_c}$
all are in the region of theoretical predictions\cite{cdf}.

Since the rest decay channels have not been observed yet, thus 
to shorten the paper, I would not list them here, but just 
to mention the most
interesting one $B_c\to J/\psi\pi$: roughly speaking,
the width is small: just about one tenth of $B_c\to J/\psi +l^+\nu_l$,
so we may expect it will be observed soon in RUN-II of Tevatron.
Since each particle of the final state in the decay can be well-detected, 
so it is the best channel to measure the mass of 
the meson $B_c$.

\vspace{2mm}
\noindent
{\bf 3 Outlooks}
\vspace{2mm}

The advances and problem(s) on double heavy mesons have been outlined.
It certainly is breakthrough progress,
that of a physical state not only the color-singlet 
components but also the color-octet 
components in Fock space Eq.(3) may play very substantial 
roles, although it still needs to be further proved. 
The progress should have wide consequences to be explored. 
Reviewing the advances on 
the topics, many subjects are just at beginning i.e. there are 
many `things' needed to be study further.

RUN-II of Tevatron will present its the first results
within two or three years. In the next run: RUN-II, 
the total statistics will be raised by one or more
order of magnitude, and the detectors will also be 
improved greatly so
the systematic errors will be suppressed and the ratio of the
signal to background will be raised much etc. In addition,
considering to LHC to be built, we believe that 
in the very near future on the color-octet mechanisms 
of heavy quarkonium production, systematic studies of the meson $B_c$
and much deeper understanding must be achieved. Namely
motivated by the foreseeing experimental achievements, including 
searching for the signatures of the color-octet proposition, 
the C.O.M.E. measurements, accumulation of more $B_c$ events 
and more accurate theoretical calculations on $B_c$ meson
etc must gain great progress soon.

One specific point which I would like to address here is that a 
comparative study of the hadronic production of the heavy
quarkonia and the meson $B_c$ is very helpful to clarify up
the situation of the color-octet proposition.
If the proposition is a correct solution for the original 
`surplus' puzzle then it is necessary
that for the production of $B_c$ meson there is no
the `surplus' at all, because the meson $B_c$ is in flavor non-singlet
that the color-singlet component contribution in the production is always 
to play a dominant role, on contrary, a heavy quarkonium 
in flavor singlet, as indicated by Figs.3:1-3, 
can gain the `surplus' by the `mixing' with a gluon
i.e. a color-octet component can be produced in a lower order of PQCD
than the color-singlet one although the former has a smaller
possibility to form
the hadron finally (C.O.M.E. is smaller).

\section*{Acknowledgments}
I would like to thank the organizers for inviting me
to talk in the conference. This work was supported partly 
by National Natural Science Foundation
of China (No.19677102, No.19910210163-A05).


\section*{References}
\begin{thebibliography}{99}
\bibitem{twk} Wu-Ki Tong, in this proceedings.

\bibitem{chang1}Chao-Hsi Chang and Yu-Qi Chen, 
\Journal{\PRD}{46}{3845}{1992}; Erratum \Journal{\PRD}{50}{6013}{1994};
Yu-Qi Chen, \Journal{\PRD}{48}{5181}{1993}; 
Erratum \Journal{\PRD}{50}{6013}{1994}.

\bibitem{bra}E. Braaten, K. Cheung and T.C. Yuan, 
\Journal{\PRD}{48}{4230}{1993}; \Journal{\PRD}{48}{5049}{1993}.

\bibitem{kis}V.V. Kiselev, A.K. Likhoded and M.V. Shevlyagin,
\Journal{\ZPC}{63}{77}{1994}; J.P. Ma,
\Journal{\PLB}{332}{398}{1994}; \Journal{\NPB}{447}{405}{1995}.

\bibitem{cdf1}CDF Collaboration, F. Abe {\it et al}, 
\Journal{\PRL}{69}{3704}{1992}; Fermilab Preprint
Fermilab-Conf-94/136-E, hep-ex/9412013.
 .
\bibitem{oct1}E. Braaten and S. Fleming, \Journal{\PRL}{74}{3327}{1995}.

\bibitem{oct2}M. Cacciari, M. Greco, M. Mangano and A. Petrelli,
\Journal{\PLB}{356}{553}{1996}; K. Cheung, W.-Y. Keung and T.C. Yuan,
\Journal{\PRL}{76}{877}{1996}; P. Cho and M. Wise,
\Journal{\PLB}{346}{129}{1995}; S. Baek, P. Ko, J. Lee and H.S. Song,
\Journal{\PRD}{55}{6839}{1997}; C.-Y. Wong and C.-W. Wong, 
\Journal{\PRD}{57}{1838}{1998}.

\bibitem{cho}P. Cho and A.K. Leibovich, \Journal{\PRD}{53}{150}{1996};
\Journal{\PRD}{53}{6203}{1996}.

\bibitem{co1}M. Beneke and M. Kr\"amer, \Journal{\PRD}{55}{R5269}{1997}.

\bibitem{co2}K. Sridhar, A.D. Martin and W.J. Stirling,
\Journal{\PLB}{438}{211}{1998}.

\bibitem{lap}G.T. Bodwin, E. Braaten and G.P. Leparge, 
\Journal{\PRD}{51}{1125}{1995}; Erratum \Journal{\PRD}{55}{5853}{1997}.

\bibitem{bch}E. Braaten and Y.-Q. Chen, \Journal{\PRL}{76}{730}{1996}.

\bibitem{chq}Chao-Hsi Chang, Cong-Feng Qiao and Jian-Xiong Wang,
\Journal{\PRD}{56}{R1363}{1997}.

\bibitem{mas}Y.-Q. Chen and Y.-P. Kuang, \Journal{\PRD}{46}{1165}{1992};
E.J. Eichten and C. Quigg, \Journal{\PRD}{49}{5845}{1994}; S.S Gershtein,
V.V. Kiselev, A.K. Likhoded and A.V. Tkabladze, 
\Journal{\PRD}{51}{3613}{1995}.

\bibitem{dec}Chao-Hsi Chang and Yu-Qi Chen, \Journal{\PRD}{49}{3399}{1994}.

\bibitem{life}I.I. Bigi, \Journal{\PLB}{371}{105}{1996}; M. Beneke and
G. Buchalla, \Journal{\PRD}{53}{4991}{1996}.

\bibitem{dec1}M. Lusignoli and M. Masetti, \Journal{\ZPC}{51}{549}{1991}.

\bibitem{prod}M. Lusignoli, M. Masetti and S. Petraren, 
\Journal{\PLB}{266}{142}{1991}; C.-H. Chang and Y.-Q.
Chen, \Journal{\PRD}{48}{4086}{1993}; C.-H. Chang and Y.-Q.
Chen, G.-P. Han and H.-T. Jiang, \Journal{\PLB}{346}{78}{1995}.

\bibitem{prod1}C.-H. Chang, Y.-Q. Chen and R.J. Oakes, 
\Journal{\PRD}{54}{4344}{1996}.

\bibitem{dec2}N. Isgur, D. Scora, B. Grinstein and M. Wise, 
\Journal{\PRD}{39}{799}{1989}; 
D. Scora and N. Isgur, \Journal{\PRD}{52}{2738}{1995};
E. Jenkins, M. Luke, A.V. Manohar and M.J. Savage, 
\Journal{\NPB}{390}{463}{1993}.

\bibitem{ins}Chao-Hsi Chang and Yu-Qi Chen, 
{\em Commun. Theor. Phys.} {\bf 23}, 451 (1995).

\bibitem{mend}S. Mandelstam, {\em Proc. R. Soc.} London {\bf 233}, 248 (1955).

\bibitem{lep}DELPHI Collaboration, P. Abreu {\it et al}, 
\Journal{\PLB}{398}{207}{1997}; OPAL Collaboration, K. Ackerstaff {\it et al}, 
\Journal{\PLB}{420}{157}{1998}; ALEPH Collaboration, R. Barac {\it et al}, 
\Journal{\PLB}{402}{213}{1997}.

\bibitem{cdf}CDF Collaboration, F. Abe {\it et al}, 
\Journal{\PRL}{81}{2432}{1998}; \Journal{\PRD}{58}{112004}{1998}.

\end{thebibliography}
\end{document}